\documentclass[preprint,12pt]{elsarticle}

\usepackage{amssymb}

\journal{Elsevier}

\begin{document}

\begin{frontmatter}



\title{SLP-Net:An efficient lightweight network for segmentation of skin lesions}
\tnotetext[mytitlenote]{We have submitted this paper to Elsevier in March 2023 and it may be published later on.}


\author[inst1]{Bo Yang}
\affiliation[inst1]{organization={School of Computer and Software Engineering},
            addressline={Xihua University}, 
            city={Chengdu},
            postcode={610039}, 
            state={Sichuan},
            country={China}}
\affiliation[inst2]{organization={School of Electrical Engineering and Electronic Information},
            addressline={Xihua University}, 
            city={Chengdu},
            postcode={610039}, 
            state={Sichuan},
            country={China}}
\affiliation[inst3]{organization={School of Computer Science},
            addressline={Nanjing University of Posts and Telecommunications}, 
            city={Nanjing},
            postcode={210023}, 
            state={Jiangshu},
            country={China}}
\author[inst1]{Hong Peng}
\author[inst1]{Chenggang Guo}
\author[inst1]{Xiaohui Luo\corref{cauthor}}
\cortext[cauthor]{Corresponding author}
\ead{xhluo_xhu@foxmail.com}
\author[inst2]{Jun Wang}
\author[inst3]{Xianzhong Long}

\begin{abstract}
Prompt treatment for melanoma is crucial. To assist physicians in identifying lesion areas precisely in a quick manner, we propose a novel skin lesion segmentation technique namely SLP-Net, an ultra-lightweight segmentation network based on the spiking neural P(SNP) systems type mechanism. Most existing convolutional neural networks achieve high segmentation accuracy while neglecting the high hardware cost. SLP-Net, on the contrary, has a very small number of parameters and a high computation speed. We design a lightweight multi-scale feature extractor without the usual encoder-decoder structure. Rather than a decoder, a feature adaptation module is designed to replace it and implement multi-scale information decoding. Experiments at the ISIC2018 challenge demonstrate that the proposed model has the highest Acc and DSC among the state-of-the-art methods, while experiments on the PH2 dataset also demonstrate a favorable generalization ability. Finally, we compare the computational complexity as well as the computational speed of the models in experiments, where SLP-Net has the highest overall superiority.
\end{abstract}



\begin{keyword}
Convolutional neural network \sep Spiking neural P systems \sep Skin lesion segmentation \sep Multi-scale feature \sep Lightweight network
\end{keyword}

\end{frontmatter}


\section{Introduction}
Melanoma is a common skin disease which, if left to develop, can develop into skin cancer and its mortality rate is quite high. Dermoscopy can be used for early lesion diagnosis and is essential for diagnosing melanoma, moreover reducing the mortality rate\cite{siegel2023cancer}. However, diagnosing dermoscopic images can be time-consuming and tedious for physicians\cite{rogers2015incidence}. In contrast, the segmentation of skin lesions with a computer is considerably more efficient. Automatic segmentation of skin lesions is considered a crucial step in the computer-aided diagnosis(CAD) of melanoma. Automatic segmentation of skin lesions in CAD is highly desirable and will greatly facilitate the dermatologist and improve the accuracy of the analysis. However, most of these systems are highly dependent on the segmentation of skin lesions to identify areas of lesions with distinct boundaries from dermoscopic images. Therefore, it is necessary to obtain accurate lesion segmentation results for dermoscopic image analysis and dermatological diagnosis.
\par Automated segmentation of skin lesions has been a topic of research for many years, and early techniques often relied on the assumption that the border of the lesion was the most essential feature for distinguishing it from the surrounding skin background. However, accurately segmenting skin lesions from healthy skin at a pixel level remains a challenging task for researchers. To address this challenge, recent research has turned to deep convolutional neural networks (DCNNs) for automated segmentation of skin lesions. DCNNs have been extensively used in medical image analysis and have achieved considerable success in various applications, including segmentation, classification, and detection of skin lesions. DCNNs have several advantages over traditional image processing techniques. They can learn features automatically from raw data and have the ability to capture complex spatial dependencies, which is particularly useful in the context of skin lesion segmentation. Moreover, DCNNs have been shown to be effective in handling noisy and irregular data, making them well-suited for the noisy and heterogeneous images often encountered in dermatology. 
\par
\begin{figure}
    \centering
    \includegraphics[width=3in]{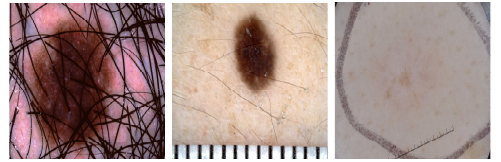}
    \caption{Images from the ISIC2018, Three common dataset segmentation challenges are demonstrated, namely, hair interference, scale interference, and lesion boundary blurring irregularities.}
    \label{fig1}
\end{figure}

 Despite the promising results of DCNN in skin lesion segmentation, there are still some challenges that need to be addressed. Firstly, deep learning algorithms tend to perform poorly when the number of samples is too small, which is a common problem in medical image analysis where datasets are often very limited. This makes it difficult to train accurate models for skin lesion segmentation. Another challenge is related to the ability of convolutional neural networks to capture information at multiple scales. Skin lesion boundaries can be highly irregular, and some images may also contain interfering factors such as hair and scale ruler as shown in Fig. \ref{fig1}. These factors can make it difficult for DCNNs to accurately segment the skin lesion from the surrounding tissue. DCNNs must be able to capture subtle differences in texture and color that distinguish skin lesions from normal skin tissue while avoiding interference from other factors. This requires the development of new segmentation techniques that can effectively extract the relevant features from skin lesion images. Besides, to the best of our knowledge, few literatures have been made to build convolutional neural networks with different neural network mechanisms. Finally, most existing segmentation convolutional neural networks such as FCN\cite{long2015fully}, U-Net\cite{ronneberger2015u}, Deeplabv3+\cite{chen2014semantic} still have a large number of parameters, which leads to longer training times and larger hardware overheads. This limits their practical application, particularly in settings where rapid diagnosis is crucial. There is a need to develop more efficient and lightweight segmentation models that can provide accurate results while minimizing computational costs. 
\par In this paper, we propose a novel convolutional neural network for skin lesion image segmentation, which is based on the spiking neural P(SNP) systems convolution mechanism. Specifically, SLP-Net is comprised of three main components, which include the SNP-type lightweight pyramid (SLP), SNP-type feature self-adaption skip connection (SFA), and SNP-type downsampling (SDS). We connect SLP in series with SDS, as shown in Fig. \ref{fig3}, to form the encoder of the network. And SFA is utilized as a decoder for each SLP separately at different resolutions. SLP is a multi-scale feature extractor, which is mainly composed of multiple sets of asymmetric convolutions with different dilated rates. As it is lightweight, it is placed in different resolution sizes for feature processing. SLP-Net is an asymmetric structure in which we discard the decoder in the network. Using a connection similar to that of ResNet\cite{he2016deep}, the proposed SFA can then effectively solve the problem about the mismatch between the encoder and the final output feature map size at all levels. SDS enables downsampling of the feature maps, thereby reducing the spatial dimensions of the data and improving computational efficiency. We conduct a comprehensive evaluation of our proposed model using two publicly available datasets ISIC2018\cite{codella2018skin} and PH2\cite{mendoncca2013ph}. We compare the performance of our model with several state-of-the-art methods in the field. The experimental results indicate that our model outperforms most existing methods in terms of accuracy while exhibiting the superior running speed and the smallest number of parameters.
\par Our main contributions are as follows:
\begin{enumerate}
    \item[1)] A novel network architecture SLP-Net is proposed, which is based on the SNP-type convolution mechanism. It has only 0.2M parameters, which is merely one hundred and fiftieth of the parameters of the classical convolutional neural network U-Net.
    \item[2)] We generalize the formulation of ConvSNP from a single channel convolutional layer to multiple channels and propose Multi-channel ConvSNP(MSConvSNP). Based on this, we propose SLP, SFA, and SDS components. They serve as feature extractors, feature adaptive jump connection methods, and downsampling, respectively, and can be ported to other networks or construct new networks with promising generality.
    \item[3)] To assess the potential of SLP-Net, we conducted a series of rigorous experiments, evaluating its performance on the publicly available datasets ISIC2018 and PH2. SLP-Net not only exhibited faster computational speed but also achieved superior segmentation performance compared to most state-of-the-art methods. 
\end{enumerate}
\section{Related work}
\subsection{Skin lesion segmentation}
Skin lesion image segmentation methods can be classified into two major categories. The first category includes classical unsupervised algorithms that rely on techniques such as edge extraction and threshold segmentation. The second category consists of modern deep learning algorithms that are based on convolutional neural networks. 
\par 
Unsupervised algorithms constitute a family of image-processing techniques that heavily rely on the extraction and recognition of low-level features that are intrinsic to the image. Remarkably, these techniques are often computationally inexpensive and don't require excessive GPU resources to operate. For example, histogram thresholding segmentation can be used to differentiate between lesioned areas in skin images and healthy ones by leveraging differences in color distribution\cite{gomez2007independent,yueksel2009accurate,mollersen2010unsupervised}. Lin et al. \cite{lin2017skin} contrasted the U-Net and C-means clustering approaches, where it was confirmed that the U-Net network segmentation results were significantly higher than the clustering approach, though the clustering approach was simple to implement and much faster to execute. An active contour model and a boundary-driven density-based clustering algorithm were also proposed by Mete et al.\cite{mete2010lesion}. To achieve faster segmentation of skin lesion areas, Peruch et al. \cite{peruch2013simpler} proposed a skin lesion image segmentation technique that integrates dimensionality reduction, color feature extraction, and post-processing. The classical methods rely on feature extraction, yielding recognition accuracy that is typically much lower than that of modern deep learning networks. This is particularly challenging when dealing with fuzzy and irregularly shaped image boundaries.
\par In recent years, deep learning has emerged as the dominant paradigm for image segmentation, detection, and recognition in computer vision. The advent of powerful GPU computing technology has facilitated the development of a plethora of deep learning convolutional neural networks (CNNs) specifically designed for medical image processing. Among these, several innovations have emerged to address specific challenges in medical image segmentation. The fully convolutional network (FCN) architecture was one of the first CNNs to achieve state-of-the-art results in semantic segmentation. Instead of fully connected layers, FCN employs convolutional layers to enable pixel-level semantic segmentation, greatly improving accuracy over traditional methods. The DeepLab\cite{chen2014semantic} architecture builds on this idea by introducing a dilated convolution operation that avoids information loss from pooling layers, and subsequently uses a conditional random field (CRF) algorithm to further optimize the segmentation process. Another influential architecture is U-Net, which symmetrizes the network structure and introduces skip connections to fuse information across different levels of semantic abstraction. U-Net has proven particularly effective in the field of medical imaging and has served as a baseline network for numerous subsequent modifications aimed at improving its segmentation ability in specific medical areas. For instance, Oktay et al.\cite{oktay2018attention} integrated the attention gating mechanism into U-Net. Tang et al.\cite{tang2019efficient} proposed a U-Net architecture with separable convolutional blocks, then addresses the overfitting problem using a random weight averaging strategy. Tang et al.\cite{tang2019multi} proposed a boundary with multi-stage U-Nets and weighted Jaccard distance loss to automatically detect lesions. Wu et al.\cite{wu2019skin} combined Inception-like convolutional blocks, recursive convolutional blocks and expanded convolutional layers to propose a modified U-Net model for segmentation of skin lesions in dermoscopic images. Benefiting from the advent of migration learning, many researchers have used models pre-trained on large datasets as feature extractors to design networks to apply them to the medical image areas. Wu et al.\cite{wu2022fat} combined the transformer structure with the U-Net encoder-decoder architecture to propose a dual-encoder segmentation network. To achieve better segmentation, Abraham et al.\cite{abraham2019novel} combine soft attention gates and U-Net, with a focal loss function based on the Tversky index to solve the problem of data imbalance in medical image segmentation. To improve the performance of U-Net in various segmentation tasks, Jha et al.\cite{jha2020doubleu} proposed DoubleU-Net with Atrous Spatial Pyramid Pooling (ASPP), which consists of two U-Net, one of which uses a pre-trained model VGG-19 as an encoder. Zhang et al.\cite{zhang2019dsm} proposed a deeply supervised multi-scale network that aggregates shallow and deep information through the utilization of the lateral output layers of the network. A multiscale connectivity block was also devised to handle the size variations of various cancers. Bi et al.\cite{bi2022hyper} proposed a hyper-fusion network to continuously optimize the pathological feature acquisition with user input features based on the separation of features from lesion images as well as user input. He et al.\cite{he2022fully} constructed a segmentation network consisting entirely of a transformer as a feature extractor. Although all the above methods obtained promising segmentation results, it was at the expense of a large amount of computational power in exchange. Furthermore, due to the presence of network pooling layers, the problem of detail and edge information loss caused by repeated downsampling is not effectively addressed.
\section{Method}
\subsection{SNP-type convolution mechanism}
\begin{figure}
    \centering
    \includegraphics[width=5in]{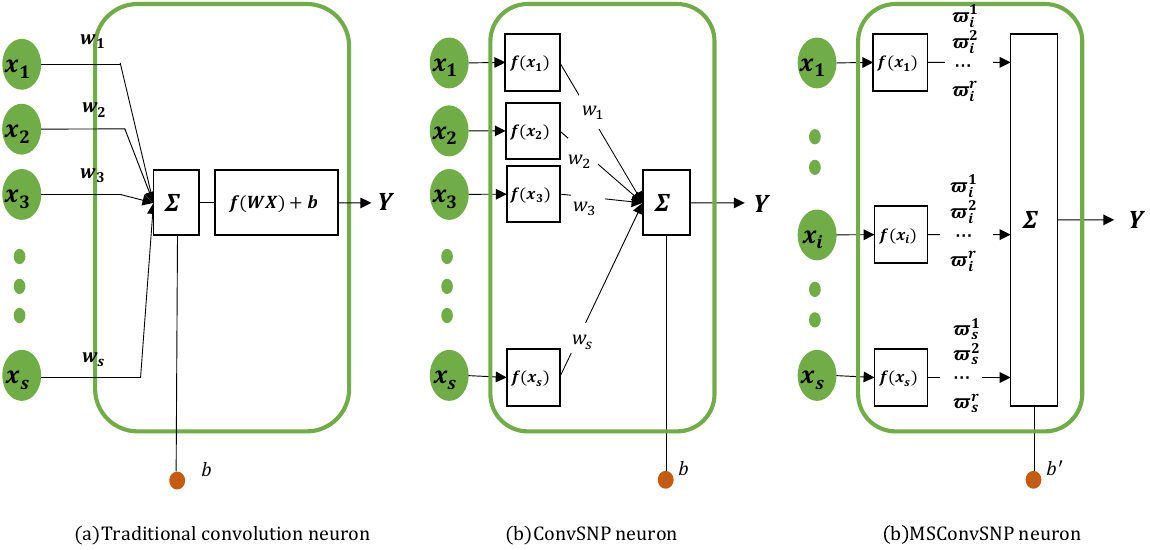}
    \caption{Three neurons with different working mechanisms}
    \label{fig2}
\end{figure}

In the past several years, SNP system and its variants have been widely applied to image processing\cite{diaz2019membrane,li2020multi,li2021medical,li2021novel,peng2021multi,mi2021medical} and time series prediction\cite{liu2021gated,liu2022lstm} as one of the classical models for membrane computing\cite{paun2010membrane}. Nonlinear spiking neural P(NSNP)\cite{peng2020nonlinear} systems is a variant of SNP systems. NSNP systems derive a deep learning model, ConvSNP\cite{zhao2022convsnp}, which has been shown to be effective when applied to networks such as ResNet, VGG, etc. because of its different working mechanism from traditional deep neural networks. Fig. \ref{fig2} shows the classical neuron and the SNP-type neuron. (a) and (b) denote conventional convolutional neurons and ConvSNP neurons, respectively. We can observe that the two types of neurons differ in their working mechanisms, mainly in the order of activation and weighted summation. For the traditional convolution neuron, let $X=\left [ x_{1}, x_{2},..., x_{s} \right ] ^{T}\in R^{s} $ denotes an input signal. As it passes through a neuron, it is multiplied and summed with the corresponding weight $W=\left [ w_{1}, w_{2},..., w_{s}\right ] ^{T}\in R^{s} $, finally biased with b as well as activated by an activation function into an output signal y. However, the signal $X$ of the ConvSNP neuron perform the activation operation before the weighted summation with $W$ to obtain output $Y$. The derivation of the formula for ConvSNP is explained very explicitly in the literature\cite{zhao2022convsnp}, please refer to it for details. The mathematical formula for the conventional convolutional neuron and the ConvSNP neuron are expressed as Eq. \ref{eq1} and Eq. \ref{eq2}, respectively.
\begin{equation}
\label{eq1}
    Y=f\left ( WX \right ) =f\left ( w_{1}x_{1}+w_{2}x_{2}+...+w_{s}x_{s}+b  \right ) 
\end{equation}
\begin{equation}
\label{eq2}
    Y=Wf\left ( X \right ) = w_{1}f(x_{1})+w_{2}f(x_{2})+...+w_{s}f(x_{s})+b
\end{equation}

where $f(\cdot)$ denotes the activation function. In extant networks, ReLU activation function is generally used. $b\in R$ indicates the bias. Further, we consider the activated input signal X having an r-channel subsignal x and set a set of weights that can be multiplied with it. The working mechanism of SNP convoluted cells can be rewritten as follows:
\begin{equation}
\label{eq2p}
    Y=Wf\left ( X \right ) = \sum_{i=1}^{r}\varpi_1^if(x_1)+\sum_{i=1}^{r}\varpi_2^if(x_2)+...+\sum_{i=1}^{r}\varpi_s^if(x_s)+b
\end{equation}
where, for any element $w$ in $W$, $w=\varpi^1+\varpi^2+... +\varpi^r,(\varpi^1,\varpi^2,...,\varpi^r)\in R^r$. The extended SNP convolutional neurons are designed based on this mechanism, we name it Multi-channel ConvSNP(MSConvSNP) which is shown in Fig. \ref{fig2}. MSConvSNP provides more scaled weight values for the input signal $X$, and it can better fit more complex data. In backpropagation, for $X$, each $\varpi$ is used individually for the solution of the gradient, instead of as a whole to find the partial derivative of loss as in Eq. \ref{eq2}. Furthermore, for the identical MSConvSNP, the weights of each scale share the same bias $b$. Therefore, only one gradient of $b$ needs to be calculated when backpropagating the gradient.

\subsection{SLP-Net architecture}
\begin{figure}
    \centering
    \includegraphics[width=5in]{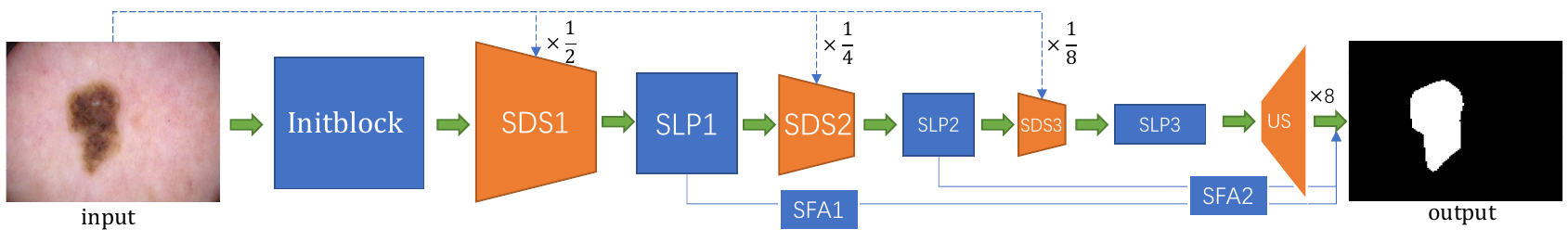}
    \caption{The  architecture of the proposed SLP-Net.}
    \label{fig3}
\end{figure}

The proposed SLP-Net is shown in Fig. \ref{fig3}. The network consists of one initblock, three SNP-type downsampling(SDS) modules, three SNP-type lightweight(SLP) modules, two SNP-type feature self-adaptation(SFA) modules, and a upsampling module(US). All SNP-type modules are built in the convolutional form of MSCovSNP.

\begin{figure}
    \centering
    \includegraphics[width=3in]{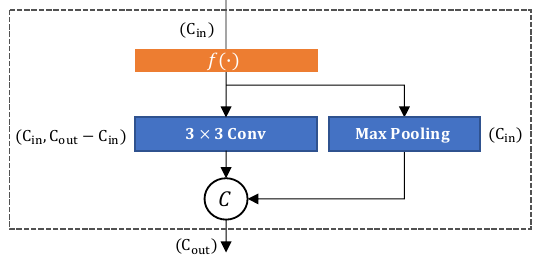}
    \caption{The proposed SDS module. The input and output channels for each layer are in parentheses. $C$ denotes concatenate.}
    \label{fig4}
\end{figure}

The initblock is a three-layer block of successive convolutional layers that converts the original three channels into sixteen channels. We use a downsampling module similar to \cite{paszke2016enet}, which concatenates a 3*3 convolution kernel and a 2*2 max-pooling, both with a step size of two. As shown in Fig. \ref{fig4}, the feature map with channel number C is downsampled by maximum pooling and convolution respectively, and the information of both downsampling operations can be utilized by concatenation. This design effectively supplements the information loss due to the pooling-only layer. By augmenting the input with additional channels of information, SLP assumes the role of a feature extractor, capable of extracting intricate details with precision. The upsampling module mainly upsamples the size of the feature map extracted by SLP3 directly eight times to the output resolution size. We arrange three SLP modules in the network to allow the model to capture information from feature maps of different resolution sizes, then add the output results together by SFA module in a ResNet manner. In the network architecture as shown in Fig. \ref{fig3}, the original image is integrated into each corresponding SDS through a process of continuous scaling. The original image is downsampled to one-half, one-quarter, and one-eighth of its original size, and subsequently concatenated with the corresponding resolution size of SDS. This supplementation of the original image information to different scale encoder requires only a small amount of computation. In the next two subsections, we will introduce the proposed SLP module and SFA module in detail.
\subsection{The proposed SLP module}
The proposed SLP module is supported by three technologies, namely, depthwise separable convolution, asymmetric convolution, and dilated convolution. In this subsection, we will briefly introduce each of the three technologies, and then gather the advantages of the three technologies to introduce the proposed SLP module.
\par \textbf{Depth-separable convolution} has been used in two prestigious models, Xception\cite{chollet2017xception} and MobileNet\cite{howard2017mobilenets}, which are two significant achievements of Google's team from the same period. Depthwise separable convolution can be divided into two types: Depthwise Convolution and Pointwise Convolution. Among them, depthwise convolution is a variant of group convolution. The number of convolution kernel channels of depthwise convolution is equal to the number of groups, that is, each input channel uses a convolution kernel separately. The standard convolution is that multiple channels share multiple convolution kernels. Furthermore, pointwise convolution can reduce dimension. Since the use of depthwise convolution has no correlation between the output map from each channel, pointwise convolution as a 1 * 1 convolution kernel can effectively increase the nonlinear relationship between the outputs of different output channels.
\par \textbf{Asymmetric convolution} is the factorization of a standard two-dimensional convolution kernel into two one-dimensional convolution kernels. In other words, an n*1 convolution kernel and a 1*n convolution kernel can be combined to form an n*n convolution kernel. This mechanism can be described in Eq. \ref{Eq3}.

\begin{equation}
\label{Eq3}
\sum_{i=-M}^{M} \sum_{j=-N}^{N} \!W(i, j) I(\alpha\! -\!i, \beta\! -\!j)\!=\! \sum_{i=-M}^{M} \!W_{\alpha }(i)[\sum_{j=-N}^{N} W_{\beta }(j) I(\alpha\! -\!i,\beta \! -\!j)]
\end{equation}
where $I$ denotes a 2D image, $W$ denotes a 2D convolution kernel, $W_\alpha$ denotes a 1D convolution kernel in the $\alpha$ dimension, and $W_\beta$ denotes a 1D convolution kernel in the $\beta$ dimension. Both $I$ and $W$ are represented in the calculation by two-dimensional matrices, $W_{\alpha}$ and $W_{\beta}$ are represented as two vectors respectively. When the 2D convolution kernel is 3*3, the decomposed convolution kernels are 1*3 and 3*1, resulting in a 33\% reduction in computational overhead from 9 to 6 parameters, though the difference in accuracy exhibited by the model is small\cite{howard2017mobilenets}.
\par \textbf{Dilated convolution} is a distinct form of convolution in which the receptive field is increased by adding zeros between two consecutive parameters in the convolution kernel, without increasing the number of parameters. The dilated convolution pyramid module is introduced in the Deeplab family\cite{chen2014semantic,chen2017deeplab,chen2018encoder} to capture information at multiple scales, which means that information is extracted from the feature map using dilated convolution with different dilation rates.
\par
\begin{figure}
    \centering
    \includegraphics[width=5in]{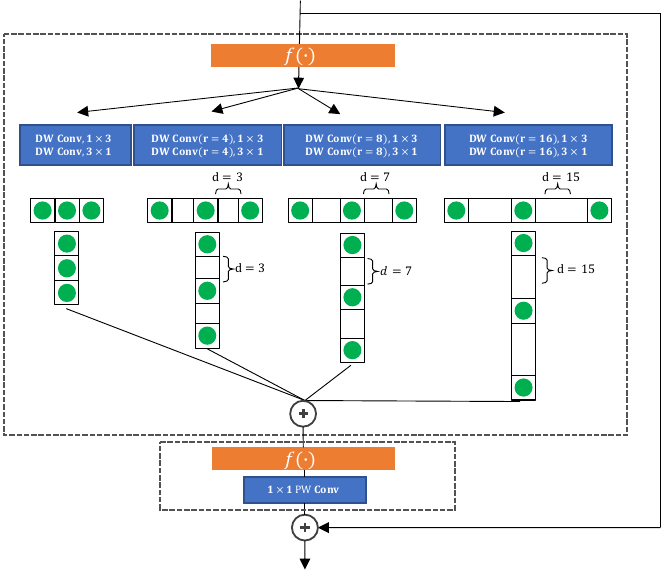}
    \caption{The proposed SLP module. DW Conv and PW Conv denote depthwise convolution and pointwise convolution, respectively. $f(\cdot)$ denotes activation function. $r$ and $d$ denote the expansion rate and the number of zeros to be filled between two neighboring parameters in the convolution kernel, respectively.}
    \label{fig5}
\end{figure}

 Inspired by the atrous spatial pyramid polling(ASPP) from the Deeplab family, we design a multi-scale feature extractor SNP-type lightweight(SLP) module, which is much more lightweight than the ASPP. The benefit of the lightweight is that we can use more SLP modules in the network, allowing it to perform feature extraction on different resolutions. As shown in Fig. \ref{fig5}, the SLP consists of two SNP-type neurons, the SNP-type depthwise convolution neuron and the SNP-type pointwise convolution neuron, both of which are indicated by dashed boxes. SNP-type depthwise convolution neuron contains four sets of convolution layers and an activation function. Each set of convolution layers is composed of 3*1 and 1*3 depthwise convolutions, whose dilated rates are 0, 4, 8, and 16, respectively. Each set of convolutions is an asymmetric convolution resulting from the deformation of a normal 3*3 convolution. The number of input channels of SLP is equal to the number of output channels. Let the number of input channels be 2N. When the feature map is convolved by the 3*1 convolution layer, the number of channels will be reduced to half of the number of 2N. The final number of output channels will be the same as the input channels after the 1*3 convolution layer. Finally, the results of several convolutions with different expansion rates are summed to obtain the feature map of 2N channels. The input image is eventually up-dimensioned to a 2N-channel feature map by a 1*1 SNP-type point convolution neuron. The SNP-type depthwise convolution neuron uses the computational mechanism of MSConvSNP, which means that the 4 groups of convolutions share a bias b and an input feature map.
The SNP-type pointwise convolution neuron serves a purpose by encoding the multiscale information received from the SNP-type depthwise convolution neuron. Its primary objective is to uncover and capture the intricate inter-relationships that exist between individual channels. The SNP-type pointwise convolution neuron and the SNP-type depthwise convolution neuron are embedded in the network in a residual-connected manner.
\subsection{The proposed SFA module}
\begin{figure}
    \centering
    \includegraphics[width=3in]{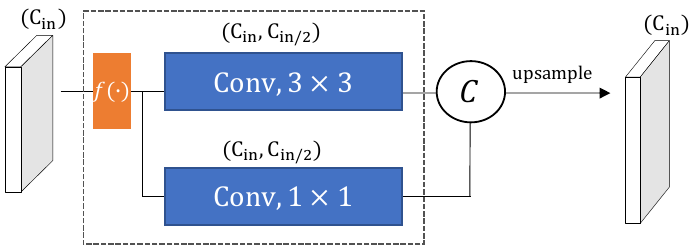}
    \caption{The proposed SFA module. DS denotes upsampling, $f(\cdot)$ denotes activation function. C denotes concatenate.}
    \label{fig6}
\end{figure}

To compensate for the loss of information caused by the feature maps after successive downsampling, we adopt skip connections like U-Net\cite{ronneberger2015u}. However, as can be seen in Fig. \ref{fig3}, the proposed network does not use a symmetrical structure similar to U-Net. We only use an 8x upsampling in the decoder stage. Therefore, when building the network, we are not able to directly concatenate the information from the encoder stage SLP to the final upsampling module. Based on the issues above, we propose the SFA module. As shown in Fig. \ref{fig6}, the SFA module is essentially an SNP-type convolution block. It should be noted that the 3*3 convolution and 1*1 convolution used together here are similar in form to the residuals. The 3*3 and 1*1 convolutions are used to adaptively encode the low-level semantic information captured by the encoder, allowing it to choose different sizes of convolution on its own. Let $X^{SFA}=\left [ x^{SFA}_{1}, x^{SFA}_{2},..., x^{SFA}_{2N} \right ] ^{T}\in R^{2N}$ be the input feature map with 2N channels, whose channel number turns to N channels after two kinds of convolution. Then both feature maps are concatenated, finally yielding the output feature map $Y^{SFA}\left [ Y^{SFA}_{1}, Y^{SFA}_{2},..., Y^{SFA}_{2N}, \right ] ^{T}\in R^{2N}$ with 2N channels. The working principle of SFA can be summarised in Eq. \ref{Eq4} and Eq. \ref{Eq5}.
\par 
\begin{equation}
\label{Eq4}
Y^{SFA}=\psi^T_3(\sigma_\lambda(X^{SFA}))\oplus\psi^T_1(\sigma_\lambda(X^{SFA}))
\end{equation}
\begin{equation}
    \label{Eq5}
    \sigma_\lambda(X^{SFA})=max(0,X^{SFA})+\lambda\times min(0,X^{SFA})
\end{equation}
where $\sigma_\lambda$ denotes PReLU activation funtion, and $\lambda$ represents the slope on the negative axis of PReLU. $\psi^T_3$ and $\psi^T_1$ denote convolutions, consisting of a set of 3*3 matrices and a set of 1*1 matrices, respectively. $\oplus$ denotes the concatenate operation. The SFA module is set up after SLP1 and SLP2. Finally, by up-sampling, we can concatenate the feature maps from various resolutions to obtain multi-scale information.
\section{Experiments}
\subsection{Datasets}
\label{sbsec4.1}
In this paper, SLP-Net is implemented and evaluated on two publicly accessible datasets.
\par The 2018 International Skin Imaging Collaboration(ISIC) skin lesion segmentation challenge dataset\cite{codella2018skin} has a total of 2594 RGB images with non-uniform resolution size. We take 2074 images from the dataset as the training set and the remaining 520 as the test set.
\par PH2 dataset\cite{mendoncca2013ph} contains 200 RGB dermoscopic images, 40 of which are melanoma and 160 benign nevus. 
\par Both datasets have the official ground truth provided.
\subsection{Implementation}
The experiments are based on the PyTorch deep learning framework. The batch size is set to 20 and the training epoch is 50. We employ Adam optimization with a learning rate of 1e-3, and a weight decay is 1e-4. Since the data set is sufficient in number, we do not increase the number of entities in the dataset. All images are scaled to 224*224 pixels and then fed into the network. It is noteworthy that, in the experiments, we have incorporated both rotated and flipped data enhancements into the dataset at the time of loading data, rather than expanding the number of entities in the dataset directly before network training. As a result, the images utilized during each epoch training round are unique, thereby enabling the network to capture and analyze the dataset from multiple perspectives while improving the accuracy and precision of the system. Also because of the data diversity it allows the network to generalize better and not overfit. To mitigate the effects of randomness and to make a fair comparison with other methods, the methods listed in the table were all trained four times and averaged in the same environment to allow for a fair comparison.
\subsection{Evaluation metrics}
To evaluate the performance of lesion segmentation,we adopt metrics recommended by the ISIC,which includes: Accuracy(Acc), Sensitivity(Sens), Specificity(Spec), Jaccard(JI) and Dice coefficient(DSC).
\begin{equation}
    Acc=\frac{TP+TN}{TP+TN+FP+FN}
\end{equation}
\begin{equation}
    Sens=\frac{Tp}{TP+FN}
\end{equation}
\begin{equation}
    Spec=\frac{TN}{TN+FP}
\end{equation}
\begin{equation}
    JI=\frac{TP}{TP+FP+FN}
\end{equation}
\begin{equation}
    DSC=\frac{2\times TP}{2\times TP+FP+FN}
\end{equation}
Where TP, TN, FP and FN are true positive, true negative, false positive and false negative respectively. All evaluation metrics have a value between 0 and 1. The closer the value is to 1, the better the segmentation result, and vice versa.
\subsection{Result}
We compare the proposed method with eight state-of-the-art methods on ISIC2018 dataset and PH2 dataset including FCN\cite{long2015fully}, SegNet\cite{badrinarayanan2017segnet}, U-Net\cite{ronneberger2015u}, AttU-Net\cite{oktay2018attention}, EDANet\cite{lo2019efficient}, PSPNet\cite{zhao2017pyramid}, Deeplabv3+\cite{chen2018encoder}, and FAT-Net\cite{wu2022fat}. PSPNet, Deeplabv3+, and FAN-Net have pre-trained models, while all other methods do not employ pre-trained models. Since the proposed SLP-Net is a lightweight architecture, in comparison with PSPNet and Deeplabv3+, two methods that also have multi-scale information extraction capabilities, we would like to use a relatively lightweight pre-trained network as their backbone to compare with our method. PSPNet and Deeplabv3+ both adopt Mobile-Net\cite{howard2017mobilenets}, which was pre-trained on Visual Object Classes Challenge 2012 (VOC2012), as their feature extractor. FAN-Net employs ResNet-34\cite{he2016deep}, pre-trained on ImageNet2012dataset, as the feature extraction model. We conduct the experiments according to the dataset partitioning approach introduced in Sec. \ref{sbsec4.1}. It is worth noting that we do not split the PH2 dataset into train and validation sets due to the small number of images in the PH2 dataset; Instead, we directly train the results derived from the 2074 images in the ISIC2018 dataset and use them for testing the PH2 dataset. The generalization ability of our model can also be checked with such an approach.
\subsection{Results on the ISIC2018 dataset.}
\begin{figure}
    \centering
    \includegraphics[width=5in]{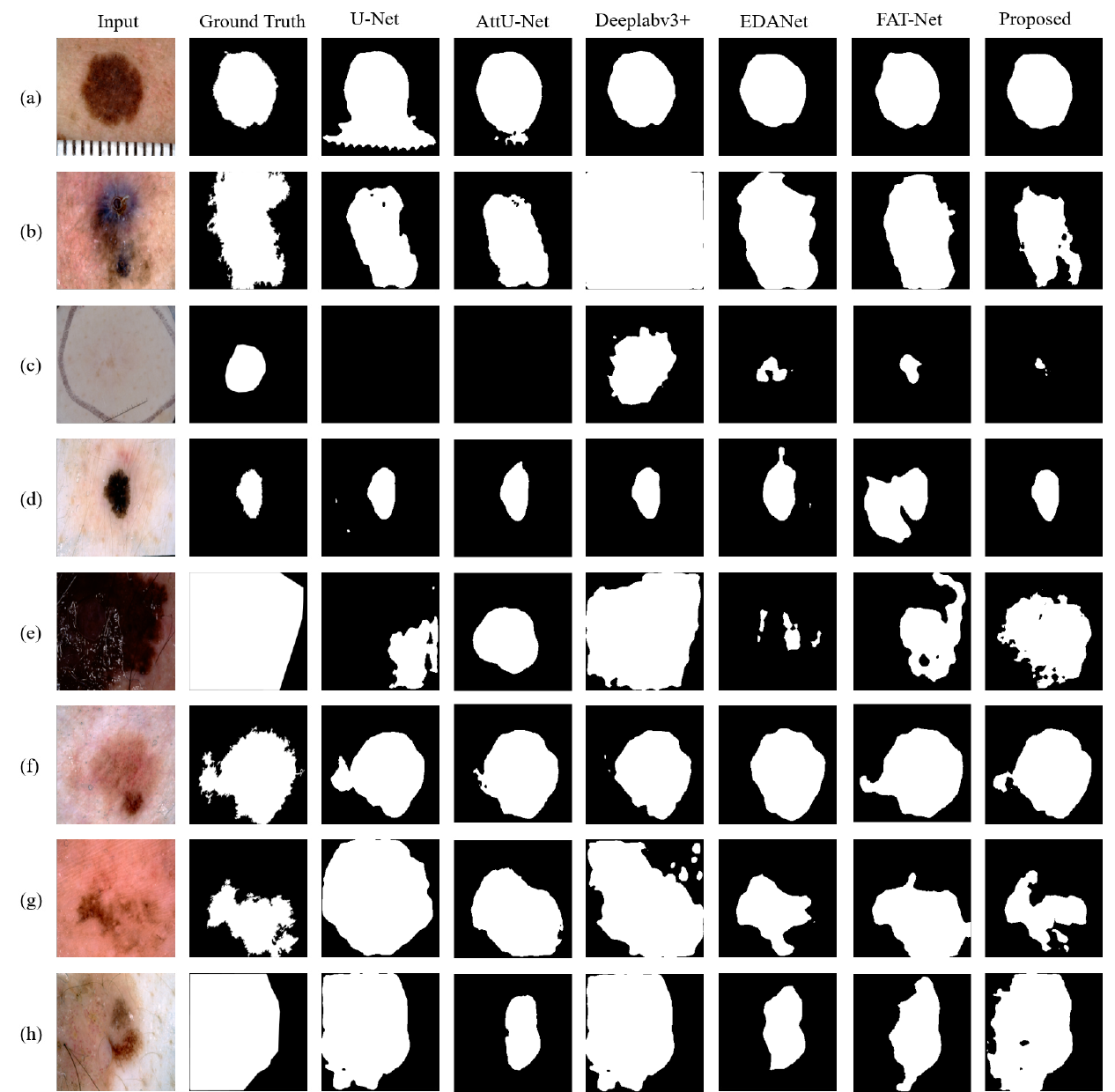}
    \caption{Segmentation results for different models on the ISIC2018 datasets. The first and second columns show the original images and the groud truth, respectively. The remaining four columns show, from left to right, the outputs of U-Net, AttU-Net, Deeplabv3+, EDANet, FAT-Net and the proposed SLP-Net.}
    \label{fig7}
\end{figure}

Tab. \ref{tab1} visualizes the metric performance of the proposed model SLP-Net and other advanced methods on the ISIC2018 dataset. The proposed SLP-Net achieved 93.87\%, 89.30\%, 95.36\%, 80.61\%, and 88.21\% in Acc, Sens, Spec, JI, and DSC metrics respectively. Among them, Acc and DSC reach the highest. U-Net, the most popular network for medical segmentation, falls behind us in terms of all metrics, except Spec. A most serious concern is that U-Net can not effectively detect images like the one illustrated in Fig. \ref{fig7}(c) where the color contrast between the lesion area and the background skin is relatively small. In other words, it has a low sensitivity. Moreover, the same problem exists in AttU-Net, a U-Net succession that has a similar structure to it. One obvious fact here is that the pre-trained networks all appear to reach higher DSC and JI metrics. However, SLP-Net, as a completely unpre-trained model, can achieve the highest DSC and the second highest JI after the first place. Deeplabv3+ achieves the highest score on Sens with 91.44\%. We fall behind him with 2.14\%. Nonetheless, Sens as a model sensitivity is required to be balanced with high DSC, otherwise, that would be erroneous, which is particularly evident in Fig. \ref{fig7}(b),(c),(g). In these three images, most of the models segments relatively small regions, whereas Deeplabv3+ incorrectly segment much larger regions. In Fig. \ref{fig7}(b), almost all of the area is segmented by Deeplabv3+ as a lesion area. 
\par
\begin{figure}
    \centering
    \includegraphics[width=5in]{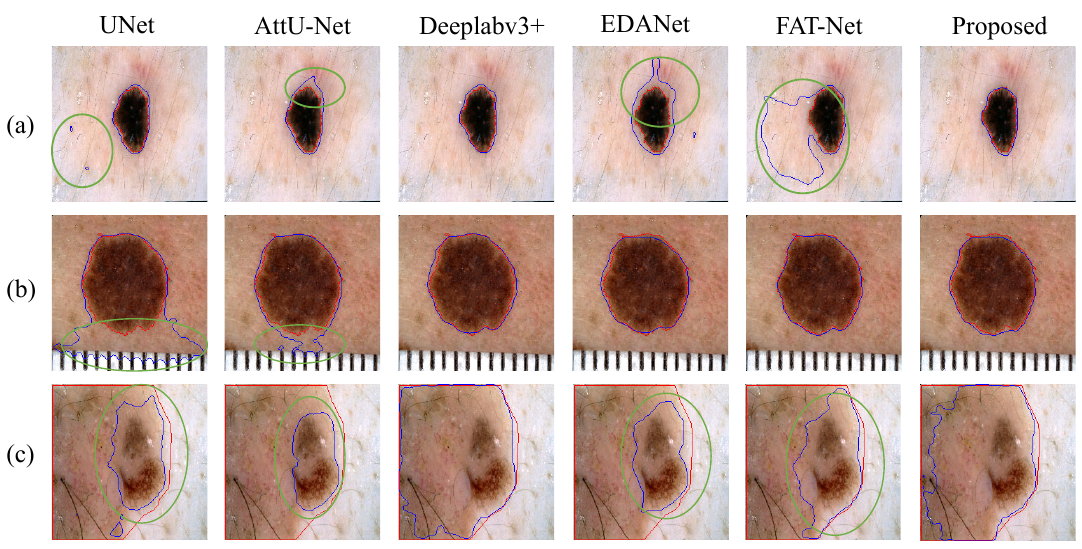}
    \caption{Comparison of segmentation details in the ISIC2018 dataset. In each figure, The parts outlined by the red and blue lines are the results of ground truth and model segmentation respectively. The area circled by the green line indicates the wrong segmentation.}
    \label{fig8}
\end{figure}

 The proposed SLP-Net can effectively detect not only the complex edges of areas of small lesions but also identify larger lesion areas, as reflected in Fig. \ref{fig8}. It is clear from row (a) that U-Net, AttU-Net, EDANet, and FAT-Net all identify non-lesioned areas as lesions with varying degrees of accuracy. We box out the misidentified areas with a green circle. Instead, due to the absence of a multi-scale feature extractor, U-Net misidentifies the scale table in the image(b) as well. Although this problem has been improved on AttU-Net, there is still redundant segmentation. (c) is also a very representative image, and many networks have difficulty identifying the lesion area. However, with the proposed SLP-Net, the lesion areas can be identified more accurately.

\begin{figure}
    \centering
    \includegraphics[width=5in]{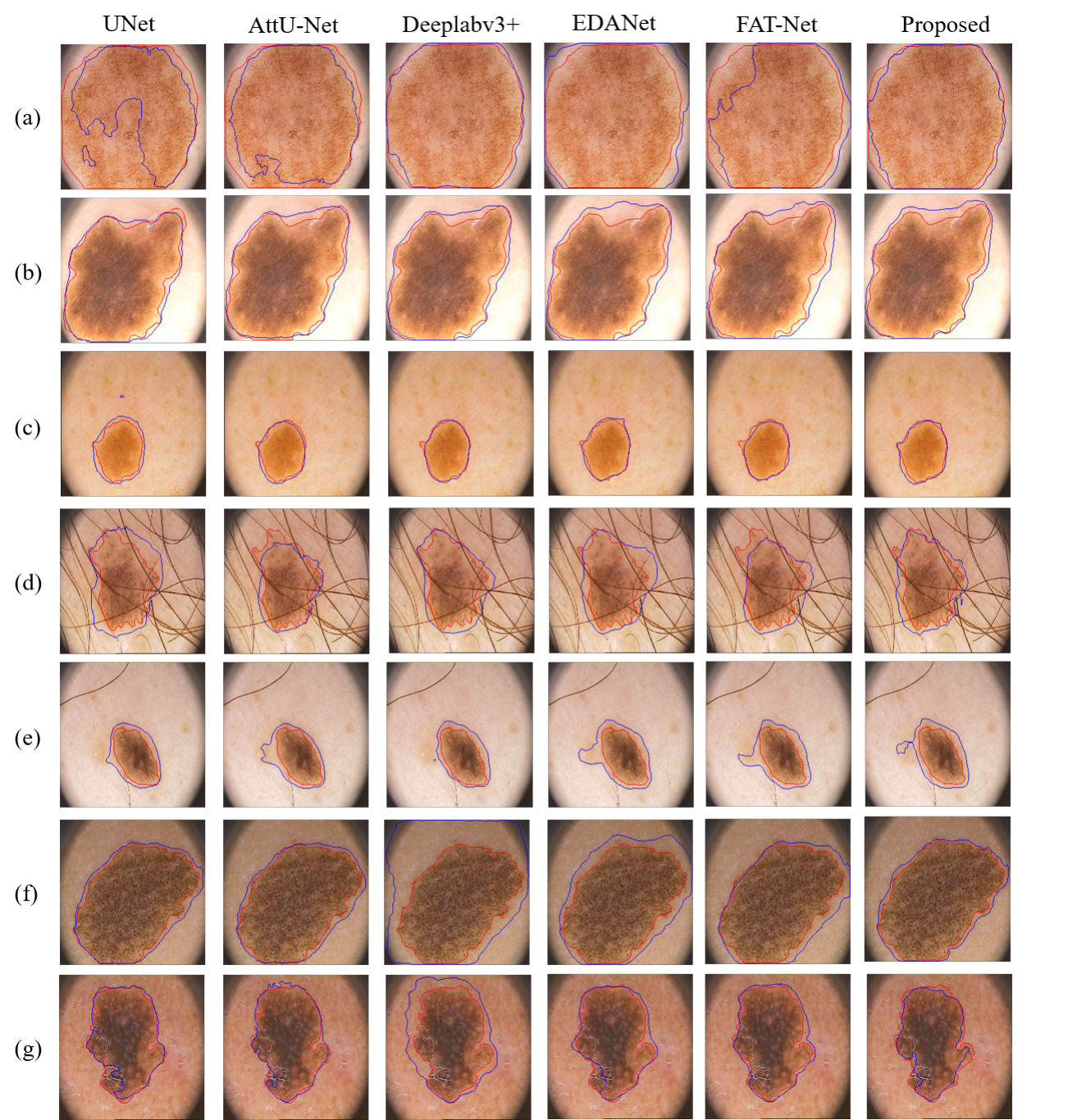}
    \caption{Segmentation results for different models on the PH2 datasets. In each figure, The parts outlined by the red and blue lines are the results of ground truth and model segmentation respectively.}
    \label{fig9}
\end{figure}
\begin{table}
    \centering
    \resizebox{\textwidth}{!}{
    \begin{tabular}{cccccc}
    \hline
    & Acc & Sens & Spec & JI & DSC  \\
    \hline
FCN8s  & 92.88 $\pm$ 0.27 & 87.19 $\pm$ 1.68 & 95.46 $\pm$ 0.95 & 78.00 $\pm$ 0.36 & 86.10 $\pm$ 0.24 \\
        SegNet & 92.11 $\pm$ 0.98 & 87.35 $\pm$ 1.76 & 94.63 $\pm$ 1.61 & 76.32 $\pm$ 2.52 & 85.08 $\pm$ 1.98 \\
        U-Net & 92.99$\pm$ 0.29 & 87.08$\pm$ 1.81 & \textbf{95.97$\pm$ 0.39} & 78.03$\pm$ 0.42 & 86.20$\pm$ 0.42 \\
         AttU-Net & 92.79$\pm$ 0.29 & 87.27$\pm$ 0.34 & 95.59$\pm$ 0.71 & 77.88$\pm$ 0.35 & 86.14$\pm$ 0.14 \\
          EDANet & 93.31$\pm$ 0.16 & 88.67$\pm$ 2.92 & 95.64$\pm$ 0.58 & 78.87$\pm$ 0.31 & 86.96$\pm$ 0.16  \\
           PSPNet(pretrained) & 93.51$\pm$ 0.12 & 89.85$\pm$ 0.96 & 93.12$\pm$ 0.47 & 79.64$\pm$ 0.24 & 87.63$\pm$ 0.14 \\
            Deeplabv3+(pretrained) & 93.67$\pm$ 0.28 & \textbf{91.44$\pm$ 0.98} & 92.72$\pm$ 1.15 & \textbf{80.66$\pm$ 0.28 }& 88.17$\pm$ 0.18 \\
            FAT-Net(pretrained) & 92.66$\pm$ 1.12 & 88.80$\pm$ 2.49 & 93.29$\pm$ 3.67 & 78.17$\pm$ 1.31 & 86.45$\pm$ 1.06 \\
            SLP-Net(Proposed) & \textbf{93.87$\pm$ 0.14} & 89.30$\pm$ 1.21 & 95.36$\pm$ 0.82 & 80.61$\pm$ 0.41 & \textbf{88.21$\pm$ 0.28 } \\
            \hline
    \end{tabular}
    }
    \caption{Results on the ISIC2018 datasets}
    \label{tab1}
\end{table}

\begin{table}
    \centering
    \resizebox{\textwidth}{!}{
    \begin{tabular}{cccccc}
    \hline
    & Acc & Sens & Spec & JI & DSC  \\
    \hline
FCN8s  & 92.88 $\pm$ 0.27 & 89.25 $\pm$ 2.86 & 93.34 $\pm$ 1.56 & 76.58 $\pm$ 1.60 & 85.64 $\pm$ 1.08 \\
        SegNet & 89.34 $\pm$ 0.75 & 89.91 $\pm$ 1.77 & 92.27 $\pm$ 1.62 & 73.36 $\pm$ 1.70 & 83.56 $\pm$ 1.30 \\
        U-Net & 90.48$\pm$ 1.29 & 88.54$\pm$ 3.85 & \textbf{94.70$\pm$ 0.90} & 76.84$\pm$ 1.95 & 85.67$\pm$ 1.74  \\
         AttU-Net & 90.48$\pm$ 1.16 & 90.67$\pm$ 0.37 & 93.67$\pm$ 1.61 & 76.62$\pm$ 3.44 & 85.78$\pm$ 2.23  \\
          EDANet & 91.11$\pm$ 0.30 & 93.75$\pm$ 1.47 & 91.18$\pm$ 1.15 & 77.45$\pm$ 0.82 & 86.45$\pm$ 0.58 \\
           PSPNet(pretrained) & 92.21$\pm$ 0.58 & 95.10$\pm$ 0.63 & 91.67$\pm$ 0.36 & 80.26$\pm$ 1.22 & 88.50$\pm$ 0.87 \\
            Deeplabv3+(pretrained) & \textbf{92.91$\pm$ 0.10} & \textbf{96.00$\pm$ 0.89} & 91.55$\pm$ 1.36 & \textbf{81.33$\pm$ 0.40 }& \textbf{89.19$\pm$ 0.35} \\
            FAT-Net(pretrained) & 92.25$\pm$ 0.13 & 93.70$\pm$ 0.13 & 93.16$\pm$ 0.13 & 81.03$\pm$ 0.25 & 88.86$\pm$ 0.17  \\
            SLP-Net(Proposed) & 91.62$\pm$ 0.60 & 92.16$\pm$ 1.06 & 93.14$\pm$ 1.69 & 79.17$\pm$ 1.78 & 87.20$\pm$ 1.25 \\
            \hline
    \end{tabular}
    }
    \caption{Results on the PH2 datasets}
    \label{tab2}
\end{table}
\subsection{Results on the PH2 dataset.}
To further validate the generalization of the model, we apply the model, pre-trained by ISIC2018, straightforwardly to the PH2 dataset for testing. The performance of the metrics on this single dataset is not satisfactory. As shown in Table \ref{tab2}, the results of the proposed method fall behind the class of pre-trained models like Deeplabv3+, PSPNet, and FAN-Net in terms of metrics. However, while the pre-trained models inherently has stronger generalization, SLP-Net also has advantages that the pre-trained model does not have, which are less training time, lower hardware consumption, and faster running speed. This is explained in detail in Sec. \ref{sbsec4.7} Then there are comparisons with the models unpre-trained, where the proposed model achieves the highest DSC and JI of the models compared in the table.
\par In the PH2 test, U-Net, AttU-Net, and FAT-Net all show excessive under-segmentation when processing image Fig. \ref{fig9}(a), in contrast to EDANet, which showed over-segmentation. Deeplabv3+ has been shown to over-segment lesion areas on the ISIC2018 dataset, with the same effect on PH2, which shows in both Fig. \ref{fig9}(f)(g). In contrast, SLP-Net shows excellent stability in the segmentation of both small lesion areas and large lesion areas without excessive under-segmentation or over-segmentation.
\begin{table}
    \centering
    \resizebox{0.6\textwidth}{!}{
    \begin{tabular}{c|cccc}
    \hline
         & GFLOPs & Params  & FPS & Params-size \\
         \hline
       FCN  & 15.45 & 15.12M & 155 & 57.67MB  \\
       SegNet  & 30.07 & 29.44M & 108 &112.32MB  \\
       U-Net & 41.90 & 31.04M & 47 &118.40MB \\
       AttU-Net & 51.01 & 34.88M & 38 &135.27MB \\
       EDA-Net & 0.85 & 0.68M & 90  & 2.60MB \\
       PSPNet(pretrained) & \textbf{0.58} & 2.41M & 115 & 9.20MB  \\
       Deeplabv3+(pretrained) & 5.05 & 5.81M & 101  &22.18MB \\
       FAT-Net(pretrained) & 30.51 & 28.76M & 36 & 123.30MB  \\
       SLP-Net & 2.30 & \textbf{0.20M} & \textbf{190} & \textbf{0.75MB} \\
       \hline
    \end{tabular}
    }
    \caption{Comparison of the number of parameters and computational complexity of each model.}
    \label{tab3}
\end{table}
\subsection{Computational complexity}
\label{sbsec4.7}
All of the experiments are implemented on a Tesla P100 GPU card with Ubuntu 18.04. Table 3 shows a comprehensive comparison of the computational complexity of all the experimented models. As can be seen from the table, the proposed model achieves better values in all metrics except GFLOPs. The highest FPS value of the proposed network is 190 while the U-Net has only 47. The processing speed of our model is about 4 times its processing speed in the same computational environment. Models that require pre-training, Deeplabv3+, PSPNet, and FAT-Net, inherently require a high pre-training cost. Due to a large number of parameters, the segmentation results depend on the effectiveness of the pre-training. Fewer parameters mean that the model can be implemented significantly easier so that it can be deployed and applied to real-time processing areas like mobile and embedded devices.
\section{Conclusion}
In this paper, we propose MSConvSNP by generalizing the formula of ConvSNP and propose a novel SNP-type convolutional neural network SLP-Net. The architecture of SLP-Net consists of two major parts, SLP and SFA, both of which are modules inspired by the SNP-type convolution mechanism. The SNP mechanism allows better results with significantly fewer parameters. As a lightweight multi-scale feature extractor, SLP stands out as a solution to complex skin lesion segmentation problems. The relatively small number of parameters allows us easily apply it to feature maps of various resolution sizes, thus facilitating information extraction at different resolutions. The SLP learns more global features, facilitating the model to distinguish between skin lesion areas of different sizes, as well as providing accurate recognition of lesion images that do not vary significantly in contrast. Furthermore, instead of the popular encoder-decoder structure, we abandon the decoder and use SFA for adaptive feature recovery. It takes a form of residual concatenation that allows adaptive recovery of boundary information, adding more high-level information to the image. Extensive experiments are done with the ISIC2018 challenge and demonstrate that better segmentation performance can be achieved with less computation cost. Better generalization of SLP-Net is also shown on the PH2 dataset compared to unpre-trained state-of-the-art networks. In our future work, we will further promote SNP and continue to explore the application of SNP in attention mechanisms.
\section*{Declaration of competing interest}
The authors declare that they have no known competing financial interests or personal relationships that could have appeared to influence the work reported in this paper.
\section*{Data availability}
Data will be made available on request.
\section*{Acknowledgement}
 \bibliographystyle{elsarticle-num} 

\begin{thebibliography}{10}
\expandafter\ifx\csname url\endcsname\relax
  \def\url#1{\texttt{#1}}\fi
\expandafter\ifx\csname urlprefix\endcsname\relax\def\urlprefix{URL }\fi
\expandafter\ifx\csname href\endcsname\relax
  \def\href#1#2{#2} \def\path#1{#1}\fi

\bibitem{siegel2023cancer}
R.~L. Siegel, K.~D. Miller, N.~S. Wagle, A.~Jemal, Cancer statistics, 2023, CA:
  a Cancer Journal for Clinicians 73~(1) (2023) 17--48.

\bibitem{rogers2015incidence}
H.~W. Rogers, M.~A. Weinstock, S.~R. Feldman, B.~M. Coldiron, Incidence
  estimate of nonmelanoma skin cancer (keratinocyte carcinomas) in the us
  population, 2012, JAMA Dermatology 151~(10) (2015) 1081--1086.

\bibitem{long2015fully}
J.~Long, E.~Shelhamer, T.~Darrell, Fully convolutional networks for semantic
  segmentation, in: Proceedings of the IEEE Conference on Computer Vision and
  Pattern Recognition, 2015, pp. 3431--3440.

\bibitem{ronneberger2015u}
O.~Ronneberger, P.~Fischer, T.~Brox, U-net: Convolutional networks for
  biomedical image segmentation, in: Medical Image Computing and
  Computer-Assisted Intervention--MICCAI 2015: 18th International Conference,
  Munich, Germany, October 5-9, 2015, Proceedings, Part III 18, Springer, 2015,
  pp. 234--241.

\bibitem{chen2014semantic}
L.-C. Chen, G.~Papandreou, I.~Kokkinos, K.~Murphy, A.~L. Yuille, Semantic image
  segmentation with deep convolutional nets and fully connected crfs, arXiv
  preprint arXiv:1412.7062 (2014).

\bibitem{he2016deep}
K.~He, X.~Zhang, S.~Ren, J.~Sun, Deep residual learning for image recognition,
  in: Proceedings of the IEEE Conference on Computer Vision and Pattern
  Recognition, 2016, pp. 770--778.

\bibitem{codella2018skin}
N.~C. Codella, D.~Gutman, M.~E. Celebi, B.~Helba, M.~A. Marchetti, S.~W. Dusza,
  A.~Kalloo, K.~Liopyris, N.~Mishra, H.~Kittler, et~al., Skin lesion analysis
  toward melanoma detection: A challenge at the 2017 international symposium on
  biomedical imaging (isbi), hosted by the international skin imaging
  collaboration (isic), in: 2018 IEEE 15th International Symposium on
  Biomedical Imaging (ISBI 2018), IEEE, 2018, pp. 168--172.

\bibitem{mendoncca2013ph}
T.~Mendon{\c{c}}a, P.~M. Ferreira, J.~S. Marques, A.~R. Marcal, J.~Rozeira, Ph
  2-a dermoscopic image database for research and benchmarking, in: 2013 35th
  Annual International Conference of the IEEE Engineering in Medicine and
  Biology Society (EMBC), IEEE, 2013, pp. 5437--5440.

\bibitem{gomez2007independent}
D.~D. Gomez, C.~Butakoff, B.~K. Ersboll, W.~Stoecker, Independent histogram
  pursuit for segmentation of skin lesions, IEEE Transactions on Biomedical
  Engineering 55~(1) (2007) 157--161.

\bibitem{yueksel2009accurate}
M.~E. Yueksel, M.~Borlu, Accurate segmentation of dermoscopic images by image
  thresholding based on type-2 fuzzy logic, IEEE Transactions on Fuzzy Systems
  17~(4) (2009) 976--982.

\bibitem{mollersen2010unsupervised}
K.~M{\o}llersen, H.~M. Kirchesch, T.~G. Schopf, F.~Godtliebsen, Unsupervised
  segmentation for digital dermoscopic images, Skin Research and Technology
  16~(4) (2010) 401--407.

\bibitem{lin2017skin}
B.~S. Lin, K.~Michael, S.~Kalra, H.~R. Tizhoosh, Skin lesion segmentation:
  U-nets versus clustering, in: 2017 IEEE Symposium Series on Computational
  Intelligence (SSCI), IEEE, 2017, pp. 1--7.

\bibitem{mete2010lesion}
M.~Mete, N.~M. Sirakov, Lesion detection in demoscopy images with novel
  density-based and active contour approaches, in: BMC bioinformatics, Vol.~11,
  BioMed Central, 2010, pp. 1--13.

\bibitem{peruch2013simpler}
F.~Peruch, F.~Bogo, M.~Bonazza, V.-M. Cappelleri, E.~Peserico, Simpler, faster,
  more accurate melanocytic lesion segmentation through meds, IEEE Transactions
  on Biomedical Engineering 61~(2) (2013) 557--565.

\bibitem{oktay2018attention}
O.~Oktay, J.~Schlemper, L.~L. Folgoc, M.~Lee, M.~Heinrich, K.~Misawa, K.~Mori,
  S.~McDonagh, N.~Y. Hammerla, B.~Kainz, et~al., Attention u-net: Learning
  where to look for the pancreas, arXiv preprint arXiv:1804.03999 (2018).

\bibitem{tang2019efficient}
P.~Tang, Q.~Liang, X.~Yan, S.~Xiang, W.~Sun, D.~Zhang, G.~Coppola, Efficient
  skin lesion segmentation using separable-unet with stochastic weight
  averaging, Computer Methods and Programs in Biomedicine 178 (2019) 289--301.

\bibitem{tang2019multi}
Y.~Tang, F.~Yang, S.~Yuan, et~al., A multi-stage framework with context
  information fusion structure for skin lesion segmentation, in: 2019 IEEE 16th
  International Symposium on Biomedical Imaging (ISBI 2019), IEEE, 2019, pp.
  1407--1410.

\bibitem{wu2019skin}
J.~Wu, E.~Z. Chen, R.~Rong, X.~Li, D.~Xu, H.~Jiang, Skin lesion segmentation
  with c-unet, in: 2019 41st Annual International Conference of the IEEE
  Engineering in Medicine and Biology Society (EMBC), IEEE, 2019, pp.
  2785--2788.

\bibitem{wu2022fat}
H.~Wu, S.~Chen, G.~Chen, W.~Wang, B.~Lei, Z.~Wen, Fat-net: Feature adaptive
  transformers for automated skin lesion segmentation, Medical Image Analysis
  76 (2022) 102327.

\bibitem{abraham2019novel}
N.~Abraham, N.~M. Khan, A novel focal tversky loss function with improved
  attention u-net for lesion segmentation, in: 2019 IEEE 16th International
  Symposium on Biomedical Imaging (ISBI 2019), IEEE, 2019, pp. 683--687.

\bibitem{jha2020doubleu}
D.~Jha, M.~A. Riegler, D.~Johansen, P.~Halvorsen, H.~D. Johansen, Doubleu-net:
  A deep convolutional neural network for medical image segmentation, in: 2020
  IEEE 33rd International Symposium on Computer-based Medical Systems (CBMS),
  IEEE, 2020, pp. 558--564.

\bibitem{zhang2019dsm}
G.~Zhang, X.~Shen, S.~Chen, L.~Liang, Y.~Luo, J.~Yu, J.~Lu, Dsm: A deep
  supervised multi-scale network learning for skin cancer segmentation, IEEE
  Access 7 (2019) 140936--140945.

\bibitem{bi2022hyper}
L.~Bi, M.~Fulham, J.~Kim, Hyper-fusion network for semi-automatic segmentation
  of skin lesions, Medical Image Analysis 76 (2022) 102334.

\bibitem{he2022fully}
X.~He, E.-L. Tan, H.~Bi, X.~Zhang, S.~Zhao, B.~Lei, Fully transformer network
  for skin lesion analysis, Medical Image Analysis 77 (2022) 102357.

\bibitem{diaz2019membrane}
D.~D{\'\i}az-Pernil, M.~A. Guti{\'e}rrez-Naranjo, H.~Peng, Membrane computing
  and image processing: a short survey, Journal of Membrane Computing 1 (2019)
  58--73.

\bibitem{li2020multi}
B.~Li, H.~Peng, J.~Wang, X.~Huang, Multi-focus image fusion based on dynamic
  threshold neural p systems and surfacelet transform, Knowledge-Based Systems
  196 (2020) 105794.

\bibitem{li2021medical}
B.~Li, H.~Peng, X.~Luo, J.~Wang, X.~Song, M.~J. P{\'e}rez-Jim{\'e}nez,
  A.~Riscos-N{\'u}{\~n}ez, Medical image fusion method based on coupled neural
  p systems in nonsubsampled shearlet transform domain, International Journal
  of Neural Systems 31~(01) (2021) 2050050.

\bibitem{li2021novel}
B.~Li, H.~Peng, J.~Wang, A novel fusion method based on dynamic threshold
  neural p systems and nonsubsampled contourlet transform for multi-modality
  medical images, Signal Processing 178 (2021) 107793.

\bibitem{peng2021multi}
H.~Peng, B.~Li, Q.~Yang, J.~Wang, Multi-focus image fusion approach based on
  cnp systems in nsct domain, Computer Vision and Image Understanding 210
  (2021) 103228.

\bibitem{mi2021medical}
S.~Mi, L.~Zhang, H.~Peng, J.~Wang, Medical image fusion based on dtnp systems
  and laplacian pyramid, Journal of Membrane Computing 3 (2021) 284--295.

\bibitem{liu2021gated}
Q.~Liu, L.~Long, H.~Peng, J.~Wang, Q.~Yang, X.~Song, A.~Riscos-N{\'u}{\~n}ez,
  M.~J. P{\'e}rez-Jim{\'e}nez, Gated spiking neural p systems for time series
  forecasting, IEEE Transactions on Neural Networks and Learning Systems
  (2021).

\bibitem{liu2022lstm}
Q.~Liu, L.~Long, Q.~Yang, H.~Peng, J.~Wang, X.~Luo, Lstm-snp: A long short-term
  memory model inspired from spiking neural p systems, Knowledge-Based Systems
  235 (2022) 107656.

\bibitem{paun2010membrane}
G.~Paun, Membrane computing, Scholarpedia 5~(1) (2010) 9259.

\bibitem{peng2020nonlinear}
H.~Peng, Z.~Lv, B.~Li, X.~Luo, J.~Wang, X.~Song, T.~Wang, M.~J.
  P{\'e}rez-Jim{\'e}nez, A.~Riscos-N{\'u}{\~n}ez, Nonlinear spiking neural p
  systems, International Journal of Neural Systems 30~(10) (2020) 2050008.

\bibitem{zhao2022convsnp}
S.~Zhao, L.~Zhang, Z.~Liu, H.~Peng, J.~Wang, Convsnp: A deep learning model
  embedded with snp-like neurons, Journal of Membrane Computing 4~(1) (2022)
  87--95.

\bibitem{paszke2016enet}
A.~Paszke, A.~Chaurasia, S.~Kim, E.~Culurciello, Enet: A deep neural network
  architecture for real-time semantic segmentation, arXiv preprint
  arXiv:1606.02147 (2016).

\bibitem{chollet2017xception}
F.~Chollet, Xception: Deep learning with depthwise separable convolutions, in:
  Proceedings of the IEEE Conference on Computer Vision and Pattern
  Recognition, 2017, pp. 1251--1258.

\bibitem{howard2017mobilenets}
A.~G. Howard, M.~Zhu, B.~Chen, D.~Kalenichenko, W.~Wang, T.~Weyand,
  M.~Andreetto, H.~Adam, Mobilenets: Efficient convolutional neural networks
  for mobile vision applications, arXiv preprint arXiv:1704.04861 (2017).

\bibitem{chen2017deeplab}
L.-C. Chen, G.~Papandreou, I.~Kokkinos, K.~Murphy, A.~L. Yuille, Deeplab:
  Semantic image segmentation with deep convolutional nets, atrous convolution,
  and fully connected crfs, IEEE Transactions on Pattern Analysis and Machine
  Intelligence 40~(4) (2017) 834--848.

\bibitem{chen2018encoder}
L.-C. Chen, Y.~Zhu, G.~Papandreou, F.~Schroff, H.~Adam, Encoder-decoder with
  atrous separable convolution for semantic image segmentation, in: Proceedings
  of the European Conference on Computer Vision (ECCV), 2018, pp. 801--818.

\bibitem{badrinarayanan2017segnet}
V.~Badrinarayanan, A.~Kendall, R.~Cipolla, Segnet: A deep convolutional
  encoder-decoder architecture for image segmentation, IEEE Transactions on
  Pattern Analysis and Machine Intelligence 39~(12) (2017) 2481--2495.

\bibitem{lo2019efficient}
S.-Y. Lo, H.-M. Hang, S.-W. Chan, J.-J. Lin, Efficient dense modules of
  asymmetric convolution for real-time semantic segmentation, in: Proceedings
  of the ACM Multimedia Asia, 2019, pp. 1--6.

\bibitem{zhao2017pyramid}
H.~Zhao, J.~Shi, X.~Qi, X.~Wang, J.~Jia, Pyramid scene parsing network, in:
  Proceedings of the IEEE Conference on Computer Vision and Pattern
  Recognition, 2017, pp. 2881--2890.

\end{thebibliography}





\end{document}